\documentclass{pasa}%

\title[Non-parametric QU-fitting]{Removing non-physical structure in fitted Faraday rotated signals: non-parametric QU-fitting}

\author[Pratley, Johnston-Hollitt, and Gaensler]{Luke Pratley$^1$\thanks{luke.pratley@gmail.com}, Melanie Johnston-Hollitt$^{2,3}$, and Bryan M. Gaensler$^{1, 4}$
\affil{$^1$Dunlap Institute for Astronomy and Astrophysics, \\ University of Toronto, Toronto, ON M5S 3H4, Canada}%
\affil{$^2$Curtin Institute for Computation, Curtin University,\\
Kent St, Bentley, WA 6102, Australia}
\affil{$^3$International Centre for Radio Astronomy Research (ICRAR), Curtin University,\\
1 Turner Ave., Technology Park, Bentley, WA 6102, Australia}
\affil{$^4$David A. Dunlap Department of Astronomy and Astrophysics, \\ University of Toronto, Toronto, ON M5S 3H4, Canada}
}%

\jid{PASA}
\doi{10.1017/pas.\the\year.xxx}
\jyear{\the\year}

\usepackage{aas_macros}
\usepackage{hyperref} 
\usepackage{bm}
\usepackage{url}
\usepackage{bbm}
\usepackage[linesnumbered,ruled]{algorithm2e}
\usepackage{amsmath}
\hypersetup{colorlinks,citecolor=blue,linkcolor=blue,urlcolor=blue}

\DeclareMathOperator*{\argmin}{argmin}

\usepackage{graphicx}
\usepackage{grffile}
\usepackage{amsmath}
\usepackage{amsfonts}
\usepackage{bm}
\begin{document}

\begin{frontmatter}
\maketitle

\begin{abstract}
Next-generation spectro-polarimetric broadband surveys will probe cosmic magnetic fields in unprecedented detail, using the magneto-optical effect known as Faraday rotation. However, non-parametric methods such as RMCLEAN can introduce non-observable linearly polarized flux into a fitted model at negative wavelengths squared. This leads to Faraday rotation structures that are consistent with the observed data, but would be impossible or difficult to measure. We construct a convex non-parametric $QU$-fitting algorithm to constrain the flux at negative wavelengths squared to be zero. This allows the algorithm to recover structures that are limited in complexity to the observable region in wavelength squared. We verify this approach on simulated broadband data sets where we show that it has a lower root mean square error and that it can change the scientific conclusions for real observations. We advise using this prior in next-generation broadband surveys that aim to uncover complex Faraday depth structures. We provide a public Python implementation of the algorithm at \url{https://github.com/Luke-Pratley/Faraday-Dreams}.
\end{abstract}

\begin{keywords}
Astrophysical magnetism, Radio astronomy, Spectropolarimetry
\end{keywords}
\end{frontmatter}

\section{Introduction} \label{sec:intro}
Faraday rotation provides a mechanism for probing magnetic fields in both the nearby and distant Universe on a range of physical and spatial scales \citep{mjh}. In Faraday rotation the angle of linearly polarized light rotates as a function of wavelength as it passes through an ionic magnetized medium. By using polarized sources as backlights, we can constrain magnetic fields in a host of environments including the interstellar and intracluster media, and potentially even the elusive cosmic web \citep{mjh}. Faraday rotation is thus vital to understanding the role of magnetic fields in the Universe.

Multiple methods have been developed to characterize the frequency dependent structure seen in spectro-polarimetric observations and thereby extract information on Faraday rotation, with the most popular approaches currently being rotation measure (RM) synthesis \citep{burn,Brentjens} and non-linear parametric fitting (\emph{i.e.} $QU$-fitting) (\emph{e.g.} \citealp{Anderson}). RMCLEAN is a CLEAN algorithm \citep{Heald} that is typically used to deconvolve the RM synthesis signal. Recent studies that use these methods to study the complexity of a Faraday rotated signal include \cite{Farnsworth_2011,OSullivan_2012,Ideguchi_2014,Kumazaki_2014,Sun_2015,Pasetto_2018,Miyashita_2018,Thomson_2021}. Each method is limited by the range and number of observed wavelengths. For previous generations of radio telescopes, observations of the emitting source have often been limited to narrow bands. However, as next-generation radio telescopes telescopes such as the Murchison Widefield Array \citep[MWA;][]{Wayth_2018,Riseley_2018,Riseley_2020}, the Low Frequency Array \citep[LOFAR;][]{vanHaarlem_2013,vanEck_2018}, the Australian Square Kilometre Array Pathfinder \citep[ASKAP;][]{Johnston_2007}, and MeerKAT \citep{Jonas_2009} observe the polarized radio sky, there is a new opportunity to constrain magnetic field models in the Universe at unprecedented precision. This has led to consideration of what method works best to determine the correct rotation measure structure from polarized spectra. With the exception of non-linear parametric $QU$-fitting, most rotation measure acquisition methods are not built for the broadband context. For example, until recently channel depolarization at low frequencies was not corrected for limiting the bands over which polarised signals could be analysed \citep{pratley}, \emph{e.g.} for telescopes such as the MWA. Broadband observations and fitting methods are needed for astronomers to have access to complex Faraday structures that are currently not either observed or understood.

In this work, we highlight a largely ignored but critical challenge when fitting broadband spectra in Faraday depth. When fitting a sinusoidal model along an axis in which we are using only measuring data collected from a region along the x-axis for which x $\geq 0$, the nature of the sinusoidal signal implies the model will also be extendable to regions that have x $\leq 0$. In the case of Faraday rotation where we are fitting flux densities in $\lambda^2$ space, where $\lambda$ is the wavelength of light, we are performing a fit over values collected for $\lambda^2 > 0$. This implies the flux density values for $\lambda^2 \leq 0$, are typically not constrained in a fitted Faraday depth model. 

However, the flux contributions for $\lambda^2 \leq 0$ can change the structures seen in the Faraday spectrum of the fitted solution. Using both simulated and real observations, we show empirically that it is possible to prevent introducing these structures in model fitting by constraining the flux to be $0$ for $\lambda^2 \leq 0$, such that the fitted model is not determined by non-observable flux at $\lambda^2 \leq 0$. We show that this suppresses structures that cannot be observed due to their fitted flux originating over $\lambda^2 \leq 0$ but otherwise will contribute to the Faraday spectrum. We emphasize that finding a $\lambda^2 \leq 0$ constrained solution has only been made possible using recent convex optimization algorithms that can include non-continuous and non-differentiable constraints and the use of RMCLEAN-like sparsity priors, \emph{e.g.}, the primal-dual based algorithm used in this work \citet{primaldual} and the alternating direction method of multipliers (ADMM) algorithm used in \citet{pratley}.

This work starts by introducing the Faraday rotation measure synthesis measurement equation in Section \ref{sec:meq}. We then discuss the aspects of flux densities for $\lambda^2 \leq 0$ and the implications in Section \ref{sec:spectro_sed}. In Section \ref{sec:algorithm} we introduce the minimisation problem that can reconstruct a Faraday rotation signal and not include the non-observable flux density in the Faraday spectrum. We demonstrate the impact of removing this flux density in signal reconstruction in Section \ref{sec:results}. We conclude that this work is important for Faraday analysis with non-parametric reconstruction algorithms like CLEAN in Section \ref{sec:conclusions}.

\section{Faraday Synthesis Measurement Equation} \label{sec:meq}
The relation between the coordinates of the Faraday spectrum, Faraday depth $\phi$, and $\lambda^2$ is given by the measurement equation
\begin{equation}
    w_{k}P(\lambda^2_k) = \int_{-\infty}^{\infty} w_{k}a(\delta\lambda^2_k, \phi)F(\phi) {\rm e}^{2i\lambda^2_k\phi} {\rm d}\phi + w_{k}n(\lambda^2_k)\, ,
\end{equation}
where $P$ is the complex valued linear polarization, $F$ is the Faraday spectrum, $n$ is the noise, $w_k$ are weights that can be used to account for uncertainty while assuming no noise co-variance, and for a limited range of $\lambda^2_k$ values and channel widths $\delta \lambda^2_k$; we are limited in both $\phi$ values and Faraday resolution $\delta \phi$ \citep{burn, Brentjens, pratley}. As discussed by \cite{pratley}, we can model the impact of channel averaging by including a channel dependent sensitivity window in Faraday depth $a(\delta\lambda^2_k, \phi)$, this is also known as the $\delta\lambda^2$-projection term and it is useful at long wavelengths. While \cite{pratley} uses channel averaging in $\lambda^2$ as an example, the averaging process is always linear by definition and different window sensitivity functions are possible  \emph{e.g.} \cite{Schnitzeler} who considered channel averaging in $\nu$. For bandlimited functions, there is an exact Fourier series relation between $\bm{y}_k = P(\lambda^2_k)$ and $\bm{x}_l = F(\phi_l)$ after including additive noise $\bm{n}_k = n(\lambda^2_k)$. We can write this relation as the matrix equation
\begin{equation}
    \bm{\mathsf{W}}\bm{y} = \bm{\mathsf{\Phi}}\bm{x} + \bm{\mathsf{W}}\bm{n}\, ,
\end{equation}
where $\bm{x}\in \mathbb{C}^N$ and $\bm{y}, \bm{n}\in \mathbb{C}^M$, and
where the measurement matrix $\bm{\mathsf{\Phi}} \in \mathbb{C}^{M \times N}$ is defined as
\begin{equation}
    \bm{\mathsf{\Phi}}_{kl} = w_{k}a(\delta\lambda^2_k, \phi_l){\rm e}^{2i\lambda^2_k\phi_l}\, ,
\end{equation}
and the diagonal weighting matrix is $\bm{\mathsf{W}}_{kk} = w_k$.
For many cases, like the examples in this paper, we can store $\bm{\mathsf{\Phi}}$ as a matrix.

\section{non-observable structure in models of broadband emission}
\label{sec:spectro_sed}

In this section, we discuss non-observable contributions of flux in the fitted model. For example, we expect that $P(\lambda^2 = 0) = 0$ due to the measured flux decreasing as $\nu \equiv c/\lambda \to \infty$. In general, the population of photons decreases to zero as energy increases\footnote{For example X-ray sources with large $\nu$ often have only 100s of photons, making the source difficult to detect in linear polarization.}. There is a more philosophical question about flux for $\lambda^2 \leq 0$. Since imaginary $i\lambda$ wavelengths do not exist, this flux corresponds to the observed Faraday rotation if it was in the opposite sense \emph{i.e.} all magnetic fields are reversed \citep{burn}. We cannot observe the energy for this signal unless this energy is shifted to positive $\lambda^2$, \emph{e.g.} through helicity \citep{Brandenburg,Horellou14}. We suggest that a reconstructed Faraday spectrum therefore should not have contributing flux over the $\lambda^2 \leq 0$ half of the domain.

Even in the case where we use conjugate symmetry to determine the flux for the negative $\lambda^2$, it is determined by the flux for positive $\lambda^2$. The negative $\lambda^2$ flux will be a factor for distinguishing and constraining Galactic magnetic field models for each line of sight component, \emph{e.g.} such modeling may be accomplished by the Interstellar MAGnetic field INference Engine (IMAGINE; \citealp{imagine}), which will perform a full Bayesian analysis of currently available polarimetric data. There are physical Faraday spectra, as suggested by \cite{Brandenburg}, which will not be consistent with conjugate symmetry in $\lambda^2 \leq 0$. This emphasizes that every model Faraday spectrum should be filtered to contain only $\lambda^2 > 0$ flux before comparing with observation. This leaves many models that are equivalent only after observable information is considered. 

Restricting to $\lambda^2 > 0$ has implications for the analysis of the Faraday spectra, which we will briefly cover. The linear polarization $P$ is related to the total intensity $I$ through
\begin{equation}
    P(\nu) = I(\nu) p(\nu)\, ,
\end{equation}
where $p(\nu)$ is the complex fractional linear polarization, and $|p| \leq 1$.\footnote{This is not always true for interferometric images.} It follows that, after a change of variables, we have the relation
\begin{equation}
    P(\lambda^2) = I(\lambda^2) p(\lambda^2)\, ,
\end{equation}
and it follows from the convolution theorem that
\begin{equation}
    F(\phi)= (K \star f)(\phi)\, .
    \label{eq:mimicry}
\end{equation}
In the above we have assumed an ideal Fourier relation with
\begin{equation}
    I(\lambda^2) = \int_{-\infty}^{\infty} K(\phi) {\rm e}^{2i\lambda^2\phi}\,  {\rm d}\phi\, .
\end{equation}
The Faraday depth coordinate $\phi$ for $K$ represents a pseudo Faraday depth component that is purely due to the spectral structure and structure of $I(\lambda^2)$ that mimics Faraday rotation modes\footnote{A smooth curve is well approximated as a slowly oscillating sine or cosine wave.} and $f$ is the Faraday rotation spectrum of $p$. We define spectral structure as structure of the spectrum in $|P(\lambda^2)|$, which is typically determined by $I(\lambda^2)$. This spectral structure over broad bandwidths creates an intrinsic broadening of the rotation measure component $\phi_0$, though we expect for many cases the resolution limit determined by the limited $\lambda^2$ coverage is far too coarse to observe this.\footnote{\citet{Brentjens} suggests working with $p(\lambda^2)$ rather than $P(\lambda^2)$ to remove spectral curvature, but be aware that $p(\lambda^2) = P(\lambda^2)/I(\lambda^2)$ has non-Gaussian distributed uncertainty. In principal if individual Faraday components have different spectral curvature this process could require more attention to detail.}

We now introduce the Heaviside step function as $\Theta(\lambda^2) = 0$ for $\lambda^2 \leq 0$ and $\Theta(\lambda^2) = 1$ otherwise. The value at $\Theta(0) = 1$ typically has no impact on its integration. The positive $\lambda^2$ linear polarization signal reads
\begin{equation}
    P_{\lambda^2 > 0}(\lambda^2) = \Theta(\lambda^2)I(\lambda^2) p(\lambda^2)\, ,
    \label{eq:step_function}
\end{equation}
with $P_{\lambda^2 \leq 0}(\lambda^2) = P(\lambda^2) -  P_{\lambda^2 > 0}(\lambda^2)$. We then have the relation
\begin{equation}
    F_{\lambda^2 > 0}(\phi) = \frac{\pi}{2i}H[F](\phi)  + \frac{1}{2}F(\phi)\, ,
     \label{eq:hilbert}
\end{equation}
where $H[F]$ is the Hilbert transform of $F(\phi)$, defined as
\begin{equation}
    H[F](\phi) = \frac{1}{\pi}\int_{-\infty}^\infty \frac{F(\phi^\prime)}{\phi - \phi^\prime}\, {\rm d}\phi^\prime\, ,
\end{equation}
which in some cases has a closed form expression. We calculate the fractional polarization without any spectral structure from total intensity $I$ as
\begin{equation}
    p_{\lambda^2 > 0}(\lambda^2) = \Theta(\lambda^2)p(\lambda^2)\, ,
\end{equation}
and
\begin{equation}
    f_{\lambda^2 > 0}(\phi) = \frac{\pi}{2i}H[f](\phi)  + \frac{1}{2}f(\phi)\, .
\end{equation}
The Hilbert transform is also encountered in all sky interferometric imaging, and there is a natural analogy between the two contexts. In Equation 27 of \citealp{pratley2}, the Hilbert transform could be used in the $uvw$-domain to restrict an all sky signal to be above the horizon. This is analogous to restricting a spectro-polarmetric signal to positive $\lambda^2$. However, unlike interferometric imaging where we can use multiple observations to get full sky coverage, we cannot build a telescope to observe and constrain negative $\lambda^2$.

\begin{figure}
    \centering
    \includegraphics[width=0.48\textwidth]{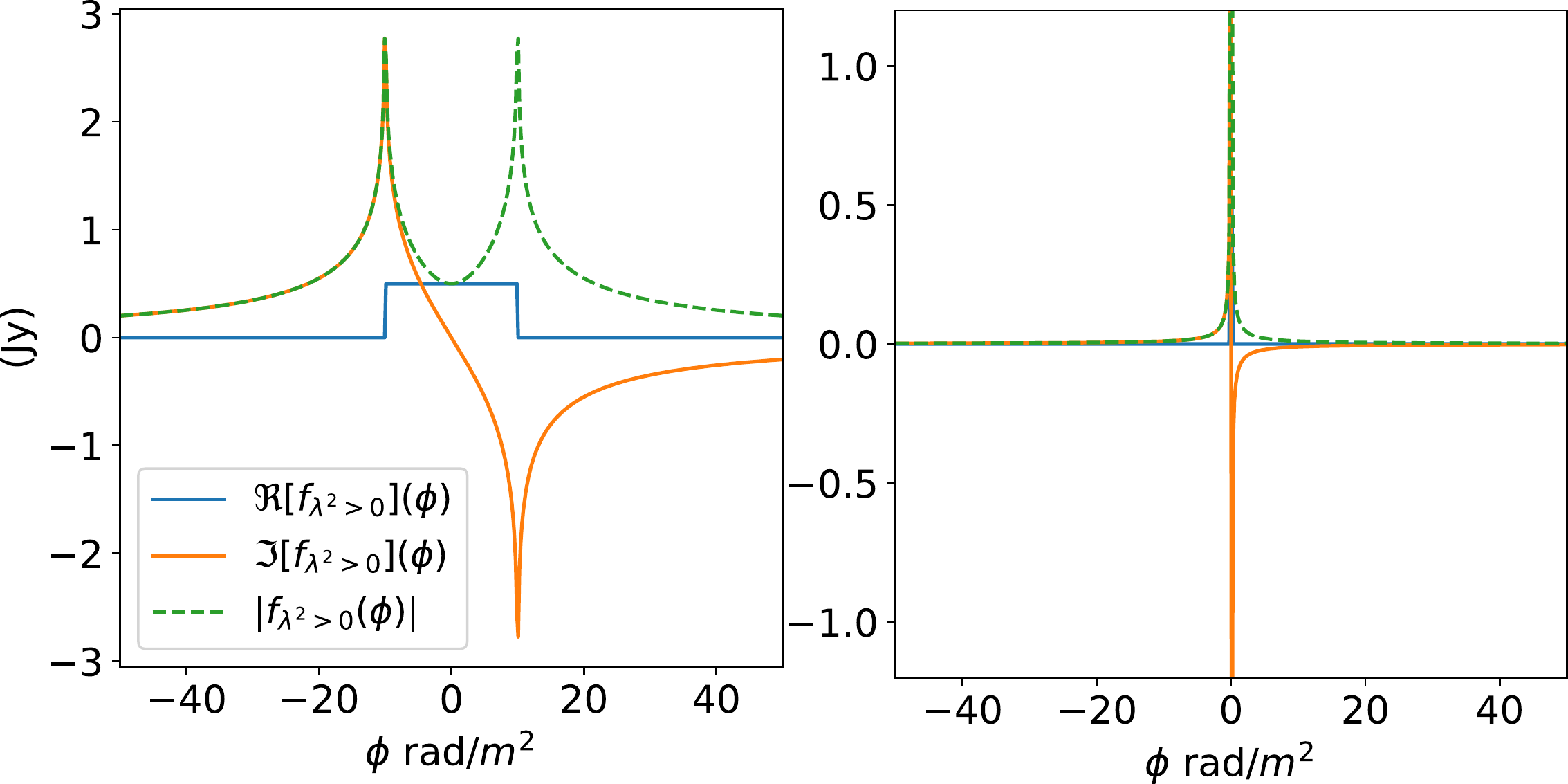}
    \caption{Real and imaginary parts for a Burn slab with the interval $\pm 10$ rad/m$^2$ (left) and for a Faraday thin screen located at $0$ rad/m$^2$ (right) in Faraday depth for $f_{\lambda^2 > 0}(\phi)$. These models only contain flux over positive $\lambda^2$. }
    \label{fig:limited_spectra}
\end{figure}

A common model for Faraday rotation with many components is the Burn slab $\Pi_{[\phi_a, \phi_b]}(\phi)$, where $\Pi_{[\phi_a, \phi_b]}(\phi) = 1$ for $\phi \in [\phi_a, \phi_b]$ and $\Pi_{[\phi_a, \phi_b]}(\phi) = 0$ otherwise \citep{burn}. The Faraday spectrum after removing non-observable structure from $\lambda^2 \leq 0$ for $f(\phi) = \Pi_{[\phi_a, \phi_b]}(\phi)$ is
\begin{equation}
   f_{\lambda^2 > 0}(\phi) = \frac{1}{2i}\ln{\left|\frac{\phi - \phi_a}{\phi - \phi_b}\right|} + \frac{1}{2}\Pi_{[\phi_a, \phi_b]}(\phi)\, .
\end{equation}
We can remove non-observable structure from $\lambda^2 \leq 0$ from a Faraday thin component $f(\phi) = \delta(\phi - \phi_0)$ centred at $\phi_0$ to read
\begin{equation}
   f_{\lambda^2 > 0}(\phi) = \frac{1}{2i (\phi - \phi_0)} + \frac{1}{2}\delta(\phi - \phi_0)\, .
\end{equation}
We can repeat the same calculations for spectral structure due to total intensity to find similar functional forms (see Figure \ref{fig:limited_spectra}).

New broadband surveys like Polarisation Sky Survey of the Universe's Magnetism (POSSUM; \citealp{possum}), Very Large Array Sky Survey (VLASS; \citealp{vlass}), and QU Observations at Cm wavelength with Km baselines using ATCA (QUOCKA\footnote{\url{https://research.csiro.au/quocka/}}) have the opportunity to measure and fit more complex Faraday structure in Faraday depth. However, the fitted complex Faraday structures should only include flux from positive $\lambda^2$. In the next section we discuss how this can be done from a fitting perspective.

\section{Non-parametric $QU$-fitting}  \label{sec:algorithm}
Recent advancements in convex optimization allow us to reconstruct non-parametric Faraday depth signals from broadband spectro-polarimetric measurements \citep{li2011,Andrecut2012,pratley,Cooray2021}. To do this, we can solve a well defined minimisation problem in the same sense that $QU$-fitting does. While $QU$-fitting is a parametric model fitting method, here we use a non-parametric model to fit $Q$ and $U$ in the Faraday dispersion spectrum with a penalty to avoid using too many components. Therefore, the method used in this work is a $QU$-fitting algorithm that fits CLEAN components \citep{Heald}, \emph{i.e.} non parametric $QU$-fitting. Both $QU$-fitting and CLEAN-style non-parametric methods are deconvolution methods and have the ability to super resolve structure. However, CLEAN does not have an explicit objective function that it will minimise. Moreover, a fitting process will make each solution consistent with the observed rotation measure synthesis signal within some error once it has been convolved to a limiting resolution. However, as we show in Section \ref{sec:results}, the flux from $\lambda^2 \leq 0$ can create structures that cannot always be removed by resolution limiting. In this work, we use a forward-backward based primal-dual algorithm \citep{primaldual} to solve the minimisation problem through a series of smaller problems without directly inverting any linear operators involved. Specifically, the primal-dual algorithm minimises both a primal problem and a dual problem at the same time. This allows each of the individual functions in the objective function to be split and minimised separately at each iteration. We use a forward-backward algorithm to minimize each individual function. One further detail of this approach is that we avoid calculating any matrix inverse or the need to perform sub-iterations, which can be computationally expensive. There are many other approaches that can solve the same minimisation problem, see \citet{Komodakis_2014} for more details. This allows us to solve the mathematical minimisation problems below, described as non-parametric $QU$-fitting problems.

In the case of Faraday thin screens, it is natural to assume that the Faraday spectrum is a sum of a few delta functions. This prior is implicitly the basis of RMCLEAN style algorithms \citep{Heald}. 
We use Bayes' theorem to relate the likelihood $\mathcal{P}(\bm{y} | \bm{x})$ and prior $\mathcal{P}(\bm{x})$ to the posterior
\begin{equation}
    \mathcal{P}(\bm{x} |\bm{y})  \propto  \mathcal{P}(\bm{y} | \bm{x})\mathcal{P}(\bm{x})\, .
\end{equation}
The solution to maximum a posteriori estimation is $\bm{x}_{\rm MAP}$, where
\begin{equation}
    \bm{x}_{\rm MAP} = \argmin_{\bm{x} \in \mathbb{C}}\left[ -\log \mathcal{P}(\bm{y} | \bm{x}) - \log \mathcal{P}(\bm{x})\right]\, .
\end{equation}
The likelihood and prior are then used to directly determine a minimisation problem. We assume that the additive noise vector $\bm{\mathsf{W}}\bm{n}$ in Equation \ref{sec:meq} follows a Gaussian distribution for both of its real and imaginary parts. The log likelihood function for a Gaussian is proportional to the squared euclidean norm that is seen in least squares minimisation. By working directly with the real and imaginary components we avoid a Rician bias when fitting the signal. We use the Laplace distribution as a sparsity prior which results in a the sum of absolute values as a penalty for the number of parameters. This gives rise to a well defined CLEAN style minimisation problem
\begin{equation}
     \bm{x}_{\rm MAP} = \argmin_{\bm{x} \in \mathbb{C}^N} \left[\gamma\|\bm{x} \|_{\ell_1} + \frac{\|\bm{\mathsf{\Phi}}\bm{x} - \bm{\mathsf{W}}\bm{y} \|_{\ell_2}^2}{2\sigma^2}\right]\ ,
\end{equation}
where the $\ell_p$-norm is defined as $\|\bm{a} \|_{\ell_p} = (\sum_k|a_k|^p)^{1/p}$. This problem is unconstrained with $\gamma$ as a parameter to be determined, and $\sigma$ is the root mean-squared (RMS) uncertainty on the measurements. This is the convex optimization problem solved by \cite{li2011} and \cite{Andrecut2012}. When the noise vector has uncorrelated components and an RMS uncertainty $\sigma_k$ for component $\bm{n}_k$, the weights are diagonal and should be chosen to be $\bm{\mathsf{W}}_{kk} = \sigma/\sigma_k$. We then choose $\sigma = \frac{1}{\sqrt{\sum_{k=1}^{M} 1/\sigma_k^2}}$ to ensure that $\sum_{k = 1}^M |\bm{\mathsf{W}}_{kk}|^2 = M$, and any further normalization is in the definition of the Fourier relation and measurement equation.

The CLEAN style prior can be recast as a constrained $\ell_1$-regularization problem
\begin{equation}
    \bm{x}_{\rm Const.} =  \argmin_{\bm{x} \in \mathbb{C}^N} \left[\|\bm{x} \|_{\ell_1} + \iota_{\mathcal{B}^\varepsilon(\bm{\mathsf{W}}\bm{y})}(\bm{\mathsf{\Phi}}\bm{x})\right]\, ,
    \label{eq:dirac_prior}
\end{equation}
where we are constraining our solution to lie close to the measurements $\bm{y}$ using an indicator function defined as $\iota_{\mathcal{U}}(\bm{a}) = 0$ when $\bm{a} \in \mathcal{U}$ and $\iota_{\mathcal{U}}(\bm{a}) = +\infty$ otherwise. The $\ell_2$-ball set $\mathcal{B}^\varepsilon(\bm{y})$ is defined as
$ \mathcal{B}^\varepsilon(\bm{\mathsf{W}}\bm{y}) = \{ \bm{z} : \|\bm{z} - \bm{\mathsf{W}}\bm{y} \|_{\ell_2} \leq \varepsilon \}$ and $\varepsilon$ is a tolerance related to $\sigma$.
The constrained formulation is closely related to the unconstrained problem and was previously used by \cite{pratley}. We can add more indicator functions or penalties as a prior to put further restrictions on the set of solutions. 

As discussed in Section \ref{sec:spectro_sed}, since no flux can be observed for $\lambda^2 \leq 0$ there is no way to constrain the fit over this range. Ignoring $P(0)$, there are special cases where the non-observable flux from $\lambda^2 \leq 0$ can be determined from observations of flux from $\lambda^2 > 0$. One special case is when $F(\phi)$ is a Hermitian function which results in the relation $P^*(-|\lambda^2|) = P(|\lambda^2|)$. However, in general, the potential to introduce structure over the range $\lambda^2 \leq 0$ is an unavoidable problem because every fitted RM component (e.g. the basis function ${\rm e}^{2i\lambda^2\phi}$) will parameterize the entire $\lambda^2$ domain. Furthermore, constraining flux to be zero for $\lambda^2 \leq 0$ limits the support of $P(\lambda^2)$, this requires the fitted signal to need more RM components and would not be promoted by a sparsity prior alone.

We can solve for the solution presented in Equations \ref{eq:step_function} and \ref{eq:hilbert} to ensure that $P(\lambda^2) \to 0$ as $\lambda^2 \to 0$ and $P(\lambda^2) = 0$ for $\lambda^2 \leq 0$, which is discussed in Section \ref{sec:spectro_sed}, so that a solution will have a Faraday spectrum that can in-principle be constrained against future observations. To do this we suggest the modification
\begin{equation}
    \bm{x}_{\rm Const.} =  \argmin_{\bm{x} \in \mathbb{C}^N} \left[\|\bm{x} \|_{\ell_1} + \iota_{\mathcal{B}^\varepsilon(\bm{\mathsf{W}}\bm{y})}(\bm{\mathsf{\Phi}}\bm{x})+ \iota_{\mathcal{C}}(\bm{\mathsf{F}}\bm{x})\right] \, ,
    \label{eq:dirac_prior_constrained}
\end{equation}
where $\bm{\mathsf{F}} \in \mathbb{C}^{N\times N}$ is a Fourier transform from $\phi$-space to $\lambda^2$-space, and $\mathcal{C} \subset \mathbb{C}^N$ is the set where $\bm{z}$ is zero for $\lambda^2 \leq 0$ for all $\bm{z}\in \mathcal{C}$. While solving Equation \ref{eq:dirac_prior_constrained}, we use Equation \ref{eq:step_function} to project onto the set of solutions that satisfy $P(\lambda^2) = P_{\lambda^2 > 0}(\lambda^2)$ within the primal-dual algorithm \citep{primaldual}. Furthermore, we can modify the constraining $\mathcal{C}$ to be resolution limited in Faraday depth, \emph{e.g.} solutions that have zero flux beyond the largest observed wavelength.; however, the impact of this appears negligible when tested by the authors.

\begin{figure*}
    \centering
    \includegraphics[width=1\textwidth]{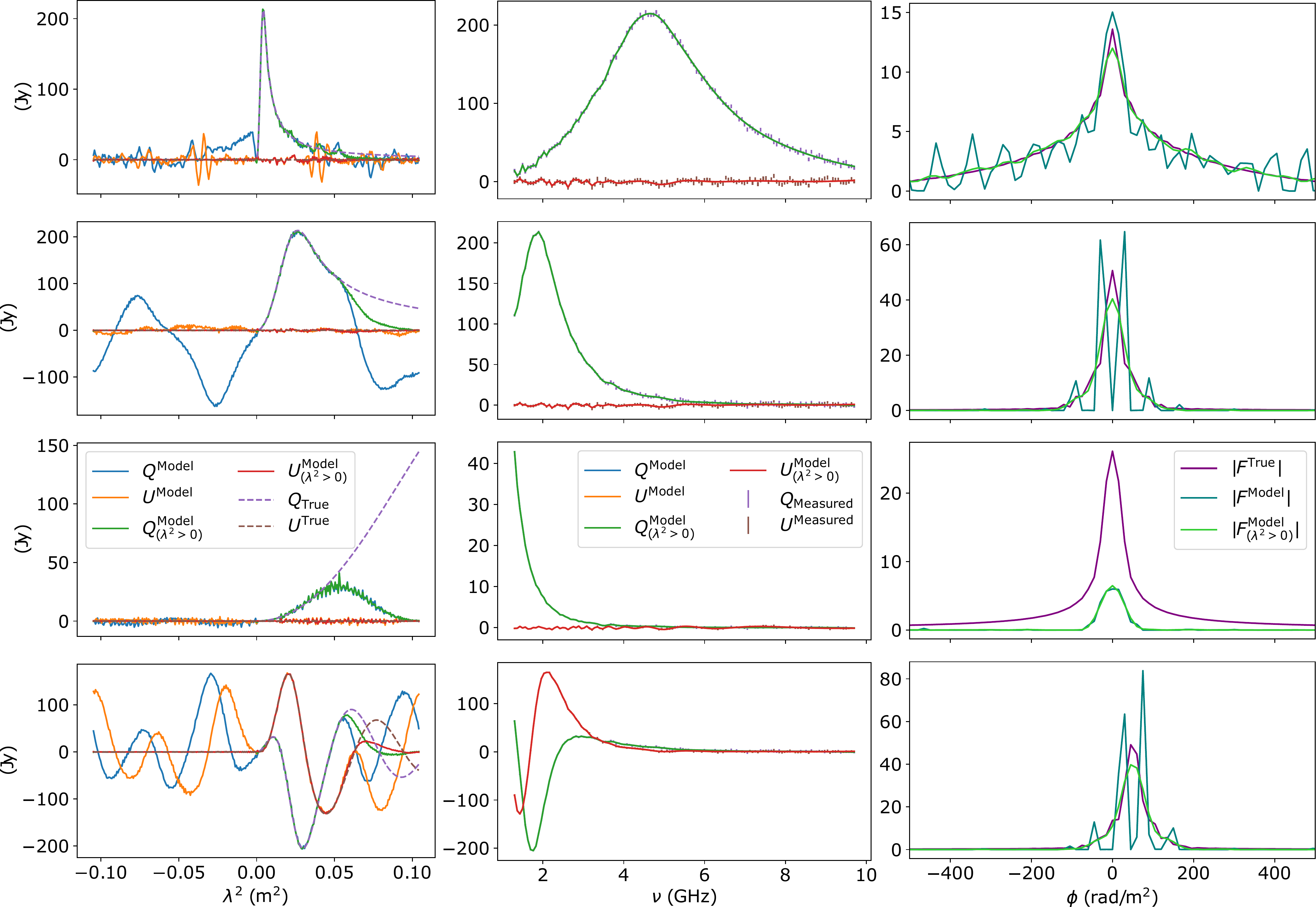}
    \caption{Reconstructions of simulated synchrotron spectra $(\nu/\nu_1)^{5/2}[1 - {\rm e}^{-(\frac{\nu}{\nu_1})^{\frac{\delta + 4}{2}}} ]$, where $\delta$ is the power law slope for the energy spectrum of cosmic-ray electrons, using a CLEAN style prior in Faraday depth, with and without the prior of $P_{\lambda^2 \leq 0}(\lambda^2) = 0$. Rows 1, 2, 3, and 4 have breaking frequencies $\nu_1 = 800, 2000, 5000, 2000$ MHz respectively with a spectral index of $\alpha \equiv -(\delta - 1)/2 = -0.8$. Rows 1 to 3 have no rotation measure, and row 4 has a component at 100 rad/m$^2$. Column 1 shows $Q(\lambda^2)$ and $U(\lambda^2)$ for the reconstructed model both with and without the prior, and the ground truth. Column 2 compares the fit over the observed $\nu$ range. Column 3 compares the absolute values of the ground truth and reconstructed Faraday depth signals. Significant spectral structure can introduce structure in $\lambda^2 \leq 0$ for the fitted signal that can never be observed. Critically, the $\lambda^2 \leq 0$ flux can significantly change the fitted model in Faraday depth, \emph{i.e.} rows 2 \& 4 where there are multiple peaks in the Faraday spectrum. Constraining $P_{\lambda^2 \leq 0}(\lambda^2) = 0$ for the fitted model removes these structures. We have also verified that this effect can be replicated for simulated spectral structure observed over frequencies $50\, {\rm MHz} \leq \nu < 1$ GHz. We also note that the shape of the ground truth Faraday spectrum $K(\phi)$ is determined by synchrotron emission (see Equation \ref{eq:mimicry}). }
    \label{fig:rm0_spectra}
\end{figure*}

\begin{figure}
    \centering
    \includegraphics[width=0.47\textwidth]{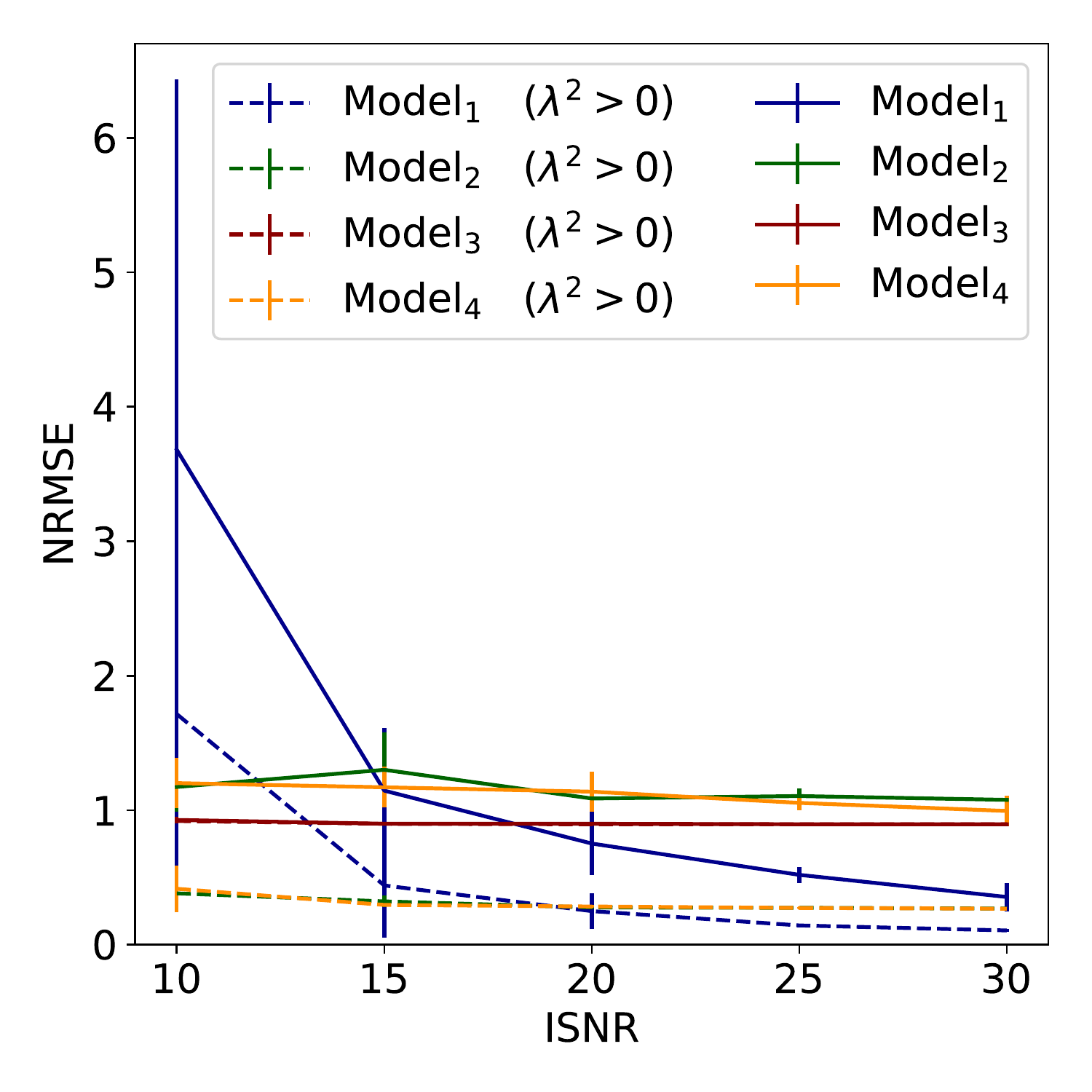}
    \caption{The NRMSE for the reconstructed Faraday spectra seen in the rows of Figure \ref{fig:rm0_spectra} (models 1 to 4 represent rows 1 to 4) for different ISNR . The error bars are centred at the mean value over 10 noise realizations and have the length given by the standard deviation.}
    \label{fig:error_plot}
\end{figure}

\begin{figure*}
    \centering
    \includegraphics[width=1\textwidth]{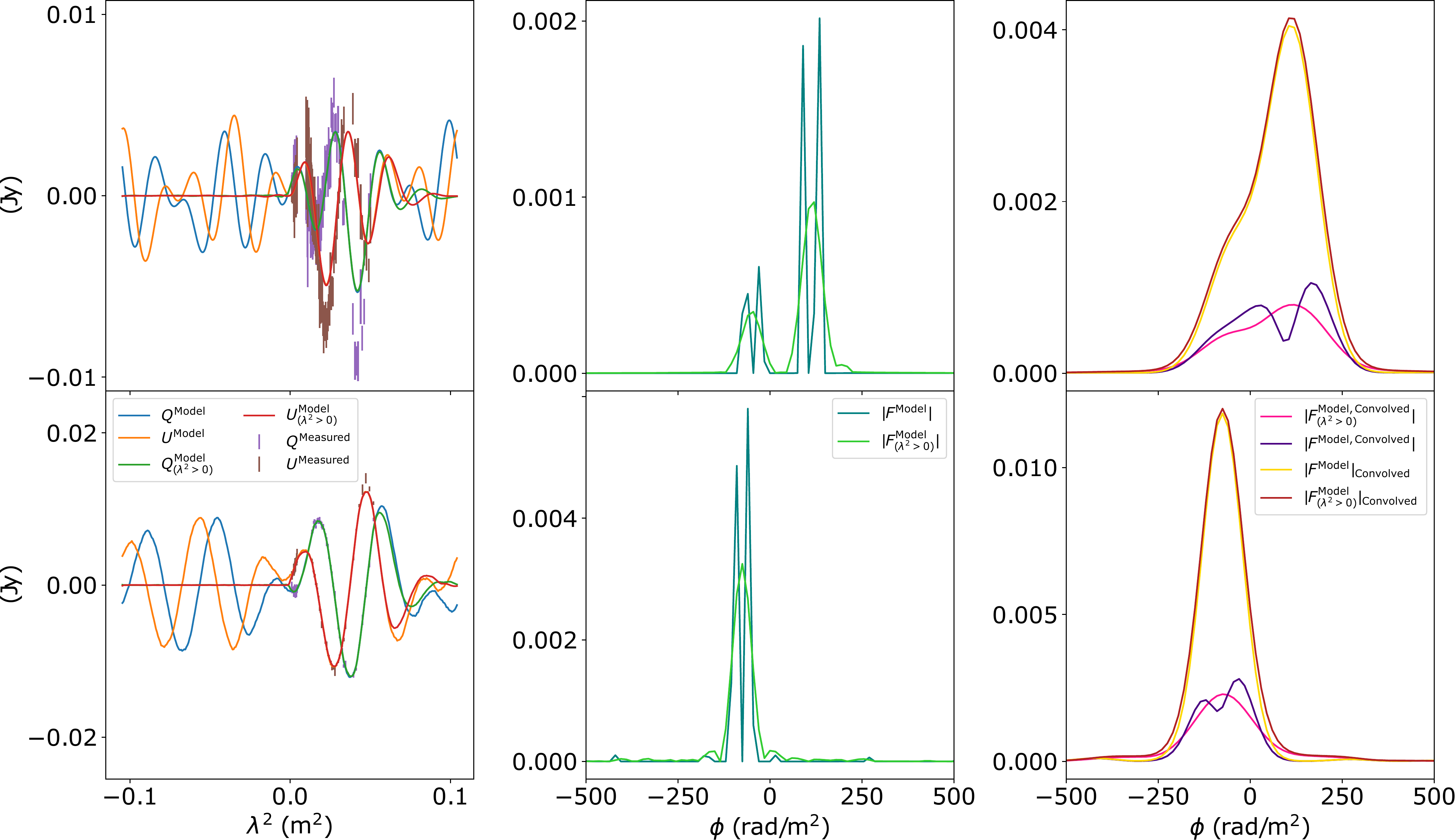}
    \caption{Reconstructions of observed spectra using a CLEAN style prior in Faraday depth with and without the prior of $P_{\lambda^2 \leq 0}(\lambda^2) = 0$, for the sources lmc\_c15 (top row) and cena\_c1972 (bottom row) from \citet{Anderson}. Columns (left to right) are measurements and fitted models in $\lambda^2$ coordinates, the fitted Faraday spectra, and the convolved Faraday spectra (where the convolutions are applied to each of the complex and absolute valued spectra).}
    \label{fig:real_data}
\end{figure*}

\begin{figure}
    \centering
    \includegraphics[width=0.47\textwidth]{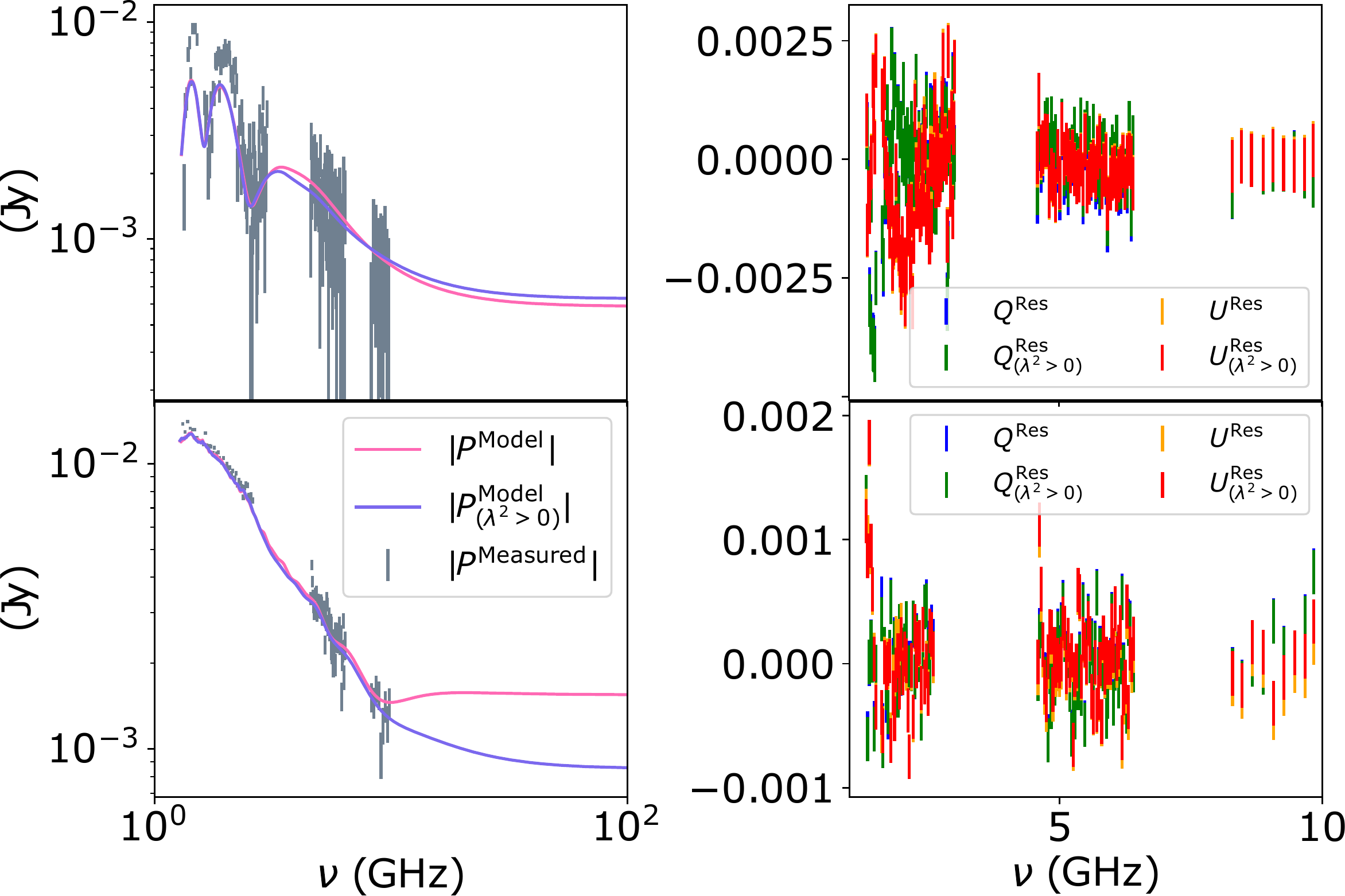}
    \caption{The magnitude and residuals for the fitted signals from Figure \ref{fig:real_data} are shown for observations of sources lmc\_c15 (top row) and cena\_c1972 (bottom row). The magnitude of the fitted linear polarization intensities and corresponding observations in $\nu$ coordinates as a logarithmic scale in the left column. The residuals for the fitted signals $Q^{\rm Res} = Q^{\rm Measured} - Q^{\rm Model}$ and $U^{\rm Res} = U^{\rm Measured} - U^{\rm Model}$ in linear scale in right column. We find that constraining $P_{\lambda^2 \leq 0}(\lambda^2) = 0$ provides a similar magnitude for the residuals, this follows because they are solutions to Equations \ref{eq:dirac_prior} and \ref{eq:dirac_prior_constrained}.}
    \label{fig:real_data_freq}
\end{figure}

\section{The simulated and observed impact of $\lambda^2 \leq 0$}
\label{sec:results}
We demonstrate that the prior for $\lambda^2 \leq 0$ can make a difference in the recovered result. One way to affect the resultant spectrum is to retain total intensity spectral structure. For a Faraday thin screen with Faraday rotation component $\phi_0$, we can write $F(\phi) = pK (\phi - \phi_0)$; here we show that the choice of prior implicitly fits a model over $\lambda^2 \leq 0$.

As a demonstration, we simulate observations of the polarization signals $P(\lambda^2) = I(\lambda^2)$ and $P(\lambda^2) = I(\lambda^2) {\rm e}^{2i\lambda^2 100}$, where $I$ is spectral structure for a synchrotron spectrum defined by Equation (5.90) of \cite{Condon}. We simulate $M = 128$ observed frequency channels that are equally spaced between 1.3 and 9.7 GHz. We then band-limit the signal by the longest wavelength in $\lambda^2$-space and use a Fourier Transform to create $\bm{x}_{\rm true}$ in Faraday depth. In interferometeric imaging, the potential reconstructed resolution is higher for large signal-to-noise ratios when using a CLEAN style prior. However, because the flux of our models is not a flat spectrum the total flux can increase or decrease with more Faraday resolution, we don't expect to accurately super-resolve the model. We follow the Nyquist resolution formula $\delta \phi \leq \frac{\pi}{2\lambda^2_{\rm max}}$; with $\lambda^2_{\rm max} = 0.0532$ we choose the resolution which is approximately twice the Nyquist sampling rate $\delta \phi = 15$ rad/m$^2$.\footnote{There is a factor of $\pi$ difference in the analogous resolution formula used in interferometry, this is due to the difference in chosen Fourier kernels \emph{e.g.} ${\rm e}^{2 i\phi \lambda^2}$ rather than ${\rm e}^{ -2i\pi\phi \lambda^2}$.} We add Gaussian noise to the Stokes $Q$ and $U$ linear polarizations individually where $P = Q + iU$, following the formula for the RMS
\begin{equation}
 \sigma = \|\bm{\mathsf{\Phi}}\bm{x}_{\rm true} \|_{\ell_2}\frac{10^{-\frac{\rm ISNR}{20}}}{\sqrt{2M}}\, ,
 \label{eq:input_noise}
\end{equation}
where ISNR is the input signal-to-noise ratio. This allows us to calculate $\varepsilon = \sqrt{2M + \sqrt{4M}}\sigma$ for ISNR $= 30$ dB. 

Figure \ref{fig:rm0_spectra} shows comparisons of the reconstructions with and without constraining $P_{\lambda^2 \leq 0}(\lambda^2) = 0$. When there is sufficient spectral structure, there is a multi-peaked structure due to flux from $\lambda^2 \leq 0$ in the solution even when there is no Faraday rotation in the signal. We also show that this is true for non-zero rotation measure values. In cases where there are no multi-peaks introduced into the Faraday spectrum, there is less unconstrained flux for $\lambda^2 \leq 0$. We show that adding the $P_{\lambda^2 \leq 0}(\lambda^2) = 0$ constraint can remove structure introduced in model fitting over the range $\lambda^2 \leq 0$. This suggests that phase information from $\lambda^2 \leq 0$ is a major contribution in this case. We calculate the normalized root mean squared error (NRMSE) between each reconstructed and ground truth Faraday spectrum using the formula ${\rm NRMSE} = \|\bm{x}_{\rm true} - \bm{x}_{\rm Const.} \|_{\ell_2}/\|\bm{x}_{\rm true} \|_{\ell_2}$. For the reconstructions and ground truths shown in column 3 of Figure \ref{fig:rm0_spectra}, the NRMSE for rows 1 to 4 for $F^{\rm Model}_{\lambda^2 > 0}(\phi)$ and $F^{\rm Model}(\phi)$ are shown in Figure \ref{fig:error_plot}. The NRMSE is lower when constraining the flux to be zero for $\lambda^2 \leq 0$, and it is comparable when the breaking frequency is not observed.

The wavelength squared range shown in Figure \ref{fig:rm0_spectra} is between the values of $\lambda^2 = \pm \frac{\pi}{2\delta \phi}$ which is $\pm 0.104$ m$^2$ when the FFT grid has a resolution of $\delta \phi = 15$ rad/m$^2$. This is the spacing of the periodic boundary conditions imposed by the Fourier series calculated using the FFT. It is important to show the full periodic range for two reasons. The first reason is that this transform can be inverted using an FFT, which means that no information is lost between the two signals. The second reason is that it is important to see that the signal wraps around at the boundaries, this phenomena is also known as aliasing. Figure \ref{fig:rm0_spectra} shows that $P_{\lambda^2 \geq 0}(\lambda^2)$ tends towards zero for large $\lambda^2$ for this reason, \emph{e.g.}, the structure of flux at large negative $\lambda^2$ will impact the structure of flux at large positive $\lambda^2$. The largest $\lambda^2$ coordinate in the model and the ground truth is ($0.104$ m$^2$) is twice the largest coordinate in the observed signal ($0.0532$ m$^2$), both of the fitted models show deviations from the ground truth spectra above $0.0532$ m$^2$.

Figure \ref{fig:real_data} demonstrates the effect and solution for two real observations from \citet{Anderson}. The sources lmc\_c15 and cena\_c1972 were observed between 1 and 10 GHz using the Australia Telescope Compact Array. The results from \cite{Anderson} show that the single component $QU$ fit seen for cena\_c1972 is consistent with the peak after the $\lambda^2 \leq 0$ correction; the three component $QU$ fit for lmc\_c15 is consistent to fitting two peaks to one component and a single peak to the other component after the $\lambda^2 \leq 0$ correction. Using non-parametric $QU$-fitting we find a smooth curve with one peak per component (lmc\_c15: 114 $\pm$ 24.5 and -50.8 $\pm$ 25.0 rad/m$^2$; cena\_c1972: -75.3 $\pm$ 28.1 rad/m$^2$), while fitting the spectral structure. These Faraday rotation measure results provided in Figure \ref{fig:real_data} are calculated by absolute flux weighting for the mean and standard deviation of Faraday depth coordinates $\phi$ with polarized flux above $0.1 \times F(\phi_{\rm peak})$. Specifically, we define the region of integration for a single component as $S = \{\phi: 0.1 \times |F(\phi_{\rm peak})| < |F(\phi)| \}$ and calculate the flux weighted mean $\langle {\rm RM} \rangle$ and standard deviation $\sigma_{{\rm RM}}$ as
\begin{equation}
    \langle {\rm RM} \rangle = \frac{\sum_{\phi \in S} |F(\phi)| \phi }{\sum_{\phi \in S} |F(\phi)| }\, ,
\end{equation}
and
\begin{equation}
    \sigma_{{\rm RM}}^2 = \frac{\sum_{\phi \in S} |F(\phi)| (\phi - \langle {\rm RM} \rangle)^2}{\sum_{\phi \in S} |F(\phi)| } \, .
 \end{equation}

Figure \ref{fig:real_data} shows that the signals with and without non-observable structure provide similar fits to the observed spectra for $\lambda^2 > 0$ in the presence of curvature. However, the structure for $\lambda^2 \leq 0$ is not constrained by the observation and imposes multiple peaks in the reconstructed signal. Figure \ref{fig:real_data_freq} shows that both constraining and unconstraining the $\lambda^2 \leq 0$ interval to zero does not greatly change the residuals in each fit, which is expected from the fidelity constraint.

In many contexts, double peaked structures can be removed by resolution limiting the signal, but this does not remove the double peaked structures for these examples. Resolution limiting the polarization magnitude $|F(\phi)|$ will remove these structures, this suggests that the phase is partially responsible for the double peaks (a similar phase issue has been discussed in \citealp{Farnsworth_2011}). However, smoothing the magnitude is a non-linear process and we do not suggest it as a method.

The reconstructions were performed using a MacBook Pro (2019) with 4 cores (2.4 GHz) and 16 GB of RAM. 50,000 iterations took approximately 15 seconds which we consider an upper bound on reconstruction time for each line of sight. However, convergence can typically be reached in 100s to 1000s of iterations which takes approximately a second or less. Each iteration applies $\bm{\mathsf{\Phi}}$ and $\bm{\mathsf{\Phi}}^\dagger$ which can be applied as either a direct matrix multiplication or using a Non-Uniform Fast Fourier transform (NUFFT). To enforce the constraint on flux in $\lambda^2$ space we need to perform an FFT and its inverse for each iteration. We can use the 15 seconds upper bound and estimate 4,200 hours of serial computation to reconstruct 1,000,000 independent lines of sight (\emph{e.g.} which we could expect in a full POSSUM catalogue; \citealp{possum}). Using a single high performance workstation with 64 cores in parallel this can be reduced to approximately an hour of computation. We provide a public Python implementation of the algorithm used in this work at \url{https://github.com/Luke-Pratley/Faraday-Dreams}.

We have shown that structures caused by fitted flux over $\lambda^2 \leq 0$ provides us with a smooth spectrum with a single peak for each Faraday screen. Without this constraint, we would arrive at a different scientific conclusion on the number of Faraday components. We have also found consistent results for the other observed broadband sources of \citet{Anderson}. 

\section{Conclusions}
\label{sec:conclusions}
We have shown that non-observable structures can be introduced into fitted models of Faraday rotation spectra when the flux for $\lambda^2 \leq 0$ is not constrained. We show that by setting the prior flux to zero over this range, we can remove the structures introduced from the unconstrained $\lambda^2 \leq 0$ region. We demonstrate the effect of this constraint using non-parametric $QU$-fitting on both simulations and real data. Without an explicit prior or constraint on $\lambda^2 \leq 0$ there can always be some contribution to the reconstructed Faraday rotation signal that is not possible to compare against future observations. This constraint will be needed when interpreting Faraday structures from next-generation broadband radio telescopes where it can impact the scientific conclusion. Current RMCLEAN algorithms do not have the ability to restrict the recovered flux only for $\lambda^2 > 0$. This will be needed in the context of interferometric observations of extended sources where fractional polarization can be non-physical, \emph{e.g.} the interstellar medium \citep{Gaensler2011}. This work shows how developments in convex optimization and polarimetric theory over the last 10 years can be leveraged for improved Faraday depth fidelity in broadband observations.

\begin{acknowledgements}
The Dunlap Institute is funded through an endowment established by the David Dunlap family and the University of Toronto. MJ-H~thanks T. Olsson, {\O}. Ber{\o}y, \& A. S{\o}rlie for motivation in the Long Night over which this manuscript was edited. BMG acknowledges the support of the Natural Sciences and Engineering Research Council of Canada (NSERC) through grant RGPIN-2015-05948, and of the Canada Research Chairs program. We thank the referee for their careful reading and constructive comments.
\end{acknowledgements}

\bibliography{sample63}

\bibliographystyle{pasa-mnras}

\end{document}